\documentclass[10pt]{article}

\usepackage{mathtools}
\usepackage{mathrsfs}
\usepackage{bm}
\usepackage{stmaryrd}
\usepackage[bbgreekl]{mathbbol}
\usepackage{hyperref}
\usepackage{subcaption}
\usepackage[margin=15pt,font=small,labelfont={bf,sf},justification=justified]{caption}
\usepackage{graphicx}
\usepackage{url}
\usepackage[per-mode=fraction]{siunitx}

\usepackage[backend=biber,bibencoding=utf8,maxnames=100
,doi=false,isbn=false,url=false,eprint=false,giveninits=true,date=year,sorting=none]{biblatex}
\renewbibmacro{in:}{}
\graphicspath{{Graphs/Results/pitch_and_rad_end}{Graphs/Results/pitch_and_rad_mid}{Graphs/Results/torque}{Graphs/Results/}{Graphs/}}

\newif\ifbembo
\newif\ifcharter
\newif\iferewhon
\newif\iflibertine
\newif\iflibertinealt
\newif\ifpalantino
\newif\iftimesnewroman

\bembofalse
\chartertrue
\erewhonfalse
\libertinefalse
\libertinealtfalse
\palantinofalse
\timesnewromanfalse

\ifbembo
  \usepackage[p,osf]{ETbb} 
  \usepackage[scaled=.95,type1]{cabin} 
  \usepackage[varqu,varl]{zi4}
  \usepackage[T1]{fontenc}
  \usepackage[libertine,vvarbb]{newtxmath}
  \usepackage[cal=boondoxo,bb=boondox]{mathalfa}
\fi

\ifcharter
  \usepackage[scaled=.98,p]{XCharter}
  \usepackage[scaled=1.04,varqu,varl]{inconsolata}
  \usepackage[type1]{cabin}
  \usepackage[uprightscript,xcharter,vvarbb,scaled=1.05]{newtxmath}
  \usepackage[cal=euler,scr=boondoxo,bb=boondox]{mathalpha}  
\fi

\iferewhon
  \usepackage[p,osf,scaled=.98,space]{erewhon} 
  \usepackage[varqu,varl]{inconsolata} 
  \usepackage[type1,scaled=.95]{cabin} 
  \usepackage[utopia,vvarbb]{newtxmath}
\fi

\iflibertine
  \usepackage[sb]{libertinus}
  \usepackage[T1]{fontenc}
  \usepackage{textcomp}
  \usepackage[varqu,varl]{zi4}
  \usepackage{libertinust1math} 
  \usepackage[scr=boondoxo,bb=boondox]{mathalpha} 
\fi

\iflibertinealt
  \usepackage[sb]{libertinus}
  \usepackage[T1]{fontenc}
  \usepackage{textcomp}
  \usepackage{libertinust1math} 
  \usepackage[scr=boondoxo,bb=boondox]{mathalpha} 
\fi

\ifpalantino
  \usepackage[largesc]{newpxtext}
  \usepackage[vvarbb]{newpxmath}
\fi

\iftimesnewroman
  \usepackage{newtxtext} 
  \usepackage[scaled=0.95]{inconsolata} 
  \usepackage[vvarbb]{newtxmath} 
\fi

\usepackage[T1]{fontenc}
\usepackage{microtype}

\usepackage{titling}
\usepackage{authblk} 
\pretitle{\begin{center}\bfseries\sffamily\LARGE}
\posttitle{\end{center}}
\usepackage{abstract}

\usepackage{sectsty} 
\allsectionsfont{\sffamily}

\usepackage[left=0.5in,right=0.5in,top=0.5in,bottom=0.5in,includefoot,heightrounded]{geometry}

\makeatletter
\patchcmd{\LS@rot}{90}{-90}{}{}
\patchcmd{\endlandscape}{90}{-90}{}{}
\makeatother

\makeatletter
\def\gsh#1{%
  \vbox{\hbox{%
    \let\\\cr
    \offinterlineskip
    \valign{&\hb@xt@2\p@{\hss$##$\hss}\vskip.2ex\cr#1\crcr}%
  }\vskip-.36ex}%
}
\def\gshsym{\@ifstar\gsh@ssym\gsh@sym}
\def\gsh@sym#1#2{\mathrlap{\overset{#1}{\phantom{#2}}}#2}
\def\gsh@ssym#1#2{\overset{#1}{#2}{\vphantom{#2}}}
\makeatother

\newcommand{\De}{\text{De}}

\renewcommand\Re{\text{Re}}

\newcommand{\uu}{\mathbf{u}}
\newcommand{\grad}{\nabla}
\newcommand{\ff}{\mathbf{f}}
\newcommand{\parens}[1]{\mathopen{}\left(#1\right)\mathclose{}}
\newcommand{\xx}{\mathbf{x}}
\newcommand{\II}{\mathbb{I}}

\newcommand{\XX}{\mathbf{X}}

\newcommand{\FF}{\mathbf{F}}

\newcommand{\bchi}{\bm{\chi}}

\newcommand{\bsigma}{\bbsigma}

\newcommand{\nn}{\mathbf{n}}

\newcommand{\CC}{\mathbb{C}}

\newcommand{\NN}{\mathbf{N}}
\newcommand{\DD}{\mathbb{D}}

\newcommand{\etal}{et al.}

\newcommand{\tran}{^{\mkern-1.5mu\mathsf{T}}}
\newcommand{\lap}{\Delta}

\newcommand{\mup}{\mu_\text{p}}
\newcommand{\mun}{\mu_\text{n}}

\newcommand{\dirVec}{\mathbf{D}}

\newcommand{\Nhigh}{N^\text{high}}
\newcommand{\Nlow}{N^\text{low}}

\title{Flagellum Pumping Efficiency in Shear-Thinning Viscoelastic Fluids}
\author[1,*]{Aaron Barrett}
\author[2]{Aaron L. Fogelson}
\author[3-6]{M. Gregory Forest}
\author[3]{Cole Gruninger}
\author[7]{Sookkyung Lim}
\author[3-6,8,9]{Boyce E. Griffith}
\affil[1]{Department of Mathematics, University of Utah, Salt Lake City, UT, USA}
\affil[2]{Departments of Mathematics and Biomedical Engineering, University of Utah, Salt Lake City, UT, USA}
\affil[3]{Department of Mathematics, University of North Carolina, Chapel Hill, NC, USA}
\affil[4]{Department of Applied Physical Sciences, University of North Carolina, Chapel Hill, NC, USA}
\affil[5]{Department of Biomedical Engineering, University of North Carolina, Chapel Hill, NC, USA}
\affil[6]{Carolina Center for Interdisciplinary Applied Mathematics, University of North Carolina, Chapel Hill, NC, USA}
\affil[7]{Department of Mathematical Sciences, University of Cincinnati, Cincinnati, Ohio, USA}
\affil[8]{Computational Medicine Program, University of North Carolina School of Medicine, Chapel Hill, NC, USA}
\affil[9]{McAllister Heart Institute, University of North Carolina School of Medicine, Chapel Hill, NC, USA}
\affil[*]{barrett@math.utah.edu}

\begin{document}

\maketitle

\begin{abstract}
Microorganism motility often takes place within complex, viscoelastic fluid environments, e.g., sperm in cervicovaginal mucus and bacteria in biofilms.  In such complex fluids, strains and stresses generated by the microorganism are stored and relax across a spectrum of length and time scales and the complex fluid can be driven out of its linear response regime.  Phenomena not possible in viscous media thereby arise from feedback between the swimmer and the complex fluid, making swimming efficiency co-dependent on the propulsion mechanism and fluid properties.  Here we parameterize a flagellar motor and filament properties together with elastic relaxation and nonlinear shear-thinning properties of the fluid in a computational immersed boundary model.  We then explore swimming efficiency, defined as a particular flow rate divided by the torque required to spin the motor, over this parameter space. Our findings indicate that motor efficiency (measured by the volumetric flow rate) can be boosted or degraded by relatively moderate or strong shear-thinning of the viscoelastic environment. 
\end{abstract}

\section{Introduction}
Microorganisms use a variety of techniques to swim in low Reynolds number environments \cite{pelczar1960,berg2003}. Organisms such as \textit{E.\ coli} use multiple rotating flagellar filaments that are connected to a rotary motor by a flexible short hook \cite{berg2003}. Spermatozoa have flagellar waveforms that are nearly planar \cite{simons2015}. Other organisms, such as \textit{Chlamydomonas reinhardtii}, use an undulating motion to propel themselves forward \cite{lauga2009}. Although the efficacy of these motions is well understood in Newtonian fluids, in Nature, these organisms frequently navigate viscoelastic environments composed of biomolecular proteins and polymeric networks solvated in a viscous Newtonian solvent. Viscoelastic dispersions induce elastic and viscous, frequency- and amplitude-dependent, responses to imposed stresses and strains. This is a completely different fluid-structure interaction system than swimming in viscous fluids in which the fluid response is instantaneous: stored elastic stress and strain in the viscoelastic fluid are released over time on spatial and temporal scales dictated by the underlying structure and organization of the macromolecular species. To avoid the expense of modeling the polymeric network at the molecular scale, the ``extra stress'' stored by the fluid's polymeric network is often modeled using continuum constitutive equations \cite{bird1987a}. One such choice is the class of upper convected Maxwell-like constitutive models such as the Oldroyd-B model \cite{larson1988,beris1994,morozov2015}, which has served as an idealized testbed for numerical method development, due to the simplicity of the material model. The Oldroyd-B model captures some features of polymeric fluids, such as generation of normal stresses in pure shear, but it fails to capture the prevalent viscoelastic property of shear-thinning. Generalizations of the Oldroyd-B model, such as the Giesekus model, were developed at the scale of polymeric kinetic theory and then coarse-grained via moment closure analysis to arrive at a constitutive law for the second-moment of the configurational probability distribution, the so-called extra-stress tensor \cite{larson1988}. A more sophisticated polymeric kinetic theory is the Rolie-Poly model for entangled polymer solutions, which along with the Giesekus model and other constitutive equations, resolve shear thinning as well as other nonlinear responses \cite{bird1987a}.

The physics of locomotion of microorgansims has been extensively studied in Newtonian fluids \cite{lauga2009}, but a complete understanding of the motion of microorganisms in a viscoelastic fluid is still lacking. There is a deep literature for non-Newtonian swimming, with results that heavily depend on both the fluid model and swimmer model. A comprehensive review of motility in non-Newtonian fluids was performed by \cite{li2021}. For undulating swimmers, \cite{lauga2007} showed that an infinite undulating sheet is always hindered by viscoelasticity. In contrast, a finite sheet can achieve a speed boost if its body shape and body elasticity is tuned to the fluid elasticity \cite{teran2010,thomases2014}. \cite{fu2009} extended the analysis to three spatial dimensions and found that generally, nonlinear viscoelasticity decreases the swimming velocity for small amplitude waves. In contrast, \cite{riley2015} showed that the swimming speed can increase provided that kinematic waves travel in opposing directions along the sheet. \cite{li2017} and \cite{hyakutake2019} found that organisms adopt a different gait in non-Newtonian fluids to enhance their swimming speed and efficiency, although Hyakutake \etal\ suggested the shear-thinning provides a greater boost than viscoelasticity. For shear thinning fluids, undulatory sheets can move more efficiently than in Newtonian counterparts \cite{velez-cordero2013,gagnon2014}, and these effects were deemed dominant over the hindrance from viscoelasticity, which is consistent with Hyakutake \etal\ In constrast, \cite{dasgupta2013} showed that there are some regimes in which viscoelasticity and shear-thinning can enhance organism motility cooperatively, depending on the fluid. Additionally, there is also evidence that continuum models do not capture the complete story and that discrete spring-dashpot models of viscoelasticity are needed to accurately understand the dynamics of swimming in these regimes \cite{wrobel2016,li2021,schuech2022}.

The literature for helical swimmers is similarly complicated. For small pitch angle infinite swimmers, viscoelasticity always results in a decrease in swimming speed \cite{fu2009,li2015a}. However, for helical swimmers with a large pitch angle, the resulting swimming speed becomes nonlinear in the Deborah number. \cite{liu2011} found that experimentally, helical swimmers in viscoelastic fluids have an enhanced swimming speed in fluids for which the relaxation time matches that of the rotation time of the helix. \cite{spagnolie2013} determined that enhanced swimming speed depends on many complex factors, such as the helical geometry, the material properties of the fluid, and the rotation rate. When considering shear thinning fluids, \cite{gomez2017} and \cite{demir2020} suggested that shear thinning results in enhanced helical swimming speeds, and that the only argument consistent with their findings is a confinement-like effect on the swimmer. Further, their results suggest that shear thinning is the dominant effect in speed enhancement because of the magnitude of speed gains over viscoelasticity. \cite{li2015a} found that the introduction of a confining cylindrical tube around a swimmer significantly enhances the swimming speed, although the details depend on the helical pitch. \cite{qu2020} determined experimentally that shear-thinning behavior is dominant over elasticity, at least in weakly elastic fluids in which the solvent shear-thins.

Herein, we determine conditions for increased versus decreased efficiency of a helical flagellum rotating in shear thinning versus non shear thinning fluids. We show that for high elasticity and low shear thinning, the swimmer can more effectively pump fluid than in a comparable Newtonian fluid. Further, we contrast our results with prior work and suggest that the underlying model for shear thinning is important. To understand the efficacy of a swimmer, one must characterize the material properties of the fluid in which they are swimming and determine an appropriate model for that particular fluid. We additionally quantify the geometric properties the swimmer takes on in these different fluids and demonstrate that the viscoelastic fluid has only mild affects on the geometry of the swimmer.

Elastic bacterial flagella can be effectively modeled using Kirchhoff rod theory because the flagellum is long ($\approx$ \SI{10}{\micro\metre}) but very thin ($\approx$ \SI{20}{\nano\meter}). Kirchhoff rod theory describes the forces and torques generated by an elastic rod in terms of the position of the centerline and an orthonormal set of director vectors attached to the center line \cite{lim2008a}. \cite{goldstein2000} introduced a bistable energy formulation for Kirchhoff rod theory to permit two stable helical configurations. \cite{darnton2007} used Kirchhoff rod theory to estimate the bending rigidity of the flagellum by fitting a Kirchhoff rod model to experimental data. \cite{lim2012} incorporated hydrodynamic interactions by creating a generalized immersed boundary method based on Kirchhoff rod theory. \cite{ko2017} incorporated a bistable energy formulation into the generalized immersed boundary method to model polymorphic transformation.

In this study, we use the generalized immersed boundary method developed by \cite{lim2012} and \cite{griffith2012} to simulate a flagellum in a shear thinning viscoelastic fluid. We assess the pumping efficiency of the flagellum and determine the shape assumed by the flagellum in different fluids.

\section{Mathematical Model}
We consider the motion of a single helical flagellum immersed in an incompressible Giesekus fluid. The fluid is described by the Cauchy stress tensor $\bsigma\parens{\xx,t}$, which consists of stress from the Newtonian solvent $\bsigma_\text{n}\parens{\xx,t}$ and stress from the embedded polymers $\bsigma_\text{p}\parens{\xx,t}$. The fluid equations of motion are
\begin{align}
\rho\frac{\partial \uu\parens{\xx,t}}{\partial t} &= \grad\cdot\bsigma\parens{\xx,t} + \ff\parens{\xx,t}, \label{eq:stokes}\\
\grad\cdot\uu\parens{\xx,t} &= 0, \\
\bsigma\parens{\xx,t} &= \bsigma_\text{n}\parens{\xx,t} + \bsigma_\text{p}\parens{\xx,t}, \\
\bsigma_\text{n}\parens{\xx,t} &= -p\parens{\xx,t}\II + \mun2\DD, \\
\bsigma_\text{p} &= \frac{\mu_\text{p}}{\lambda}\parens{\CC - \II} \\
\gshsym{\gsh{\triangledown}}{\CC}\parens{\xx,t} &= \frac{-1}{\lambda}\parens{\CC\parens{\xx,t}-\II} - \frac{\alpha}{\lambda}\parens{\CC\parens{\xx,t}-\II}^2 + D\lap\parens{\CC\parens{\xx,t} - \II}. \label{eq:conform}
\end{align}
in which $\DD = \frac{1}{2}\parens{\grad\uu\parens{\xx,t}+\grad\uu\parens{\xx,t}\tran}$ is the rate of strain tensor, $\uu\parens{\xx,t}$ is the Eulerian fluid velocity, $p\parens{\xx,t}$ is the pressure, $\ff\parens{\xx,t}$ is the external body force density acting on the fluid, $\mun$ and $\mup$ are the Newtonian solvent and polymeric contributions to the viscosity, respectively, $\lambda$ is the relaxation time of the fluid, $D$ is the diffusion coefficient of the stress, and $\gshsym{\gsh{\triangledown}}{\CC}\parens{\xx,t}$ is the upper convected derivative defined by
\begin{equation}
\gshsym{\gsh{\triangledown}}{\CC}\parens{\xx,t} = \frac{\partial \CC\parens{\xx,t}}{\partial t} + \uu\parens{\xx,t}\cdot\grad\CC\parens{\xx,t} - \parens{\CC\parens{\xx,t}\cdot\grad\uu\parens{\xx,t}\tran+\grad\uu\parens{\xx,t}\cdot\CC\parens{\xx,t}}.
\end{equation}
The parameter $\alpha$ governs the strength of the nonlinear anisotropic drag term, which we refer to as the nonlinear parameter in the remainder of this work. This parameter determines the degree to which the fluid experiences shear thinning. Large values of $\alpha$ correspond to an enhanced capacity for shear-thinning. The Giesekus model reduces to the Oldroyd-B model in the limit of $\alpha\rightarrow 0$. In the continuum equations of motion, the conformation tensor $\CC\parens{\xx,t}$ should remain positive definite at all times during the computation. If, during the course of the simulation, the conformation tensor loses positive definiteness, we project the conformation tensor onto the nearest non-negative-definite tensor \cite{guy2008}.

We note the use of the unsteady Stokes equations in \eqref{eq:stokes} as opposed to the steady Stokes equations. In our simulations, we utilize adaptive mesh refinement around the flagellum, which is precisely where we want to capture greater accuracy. However, our discretization of the Stokes operator near coarse-fine interfaces limits the solver to non-zero Reynolds numbers \cite{gruninger2024}.

It is well established that the Oldroyd-B model exhibits instabilities near extensional points \cite{renardy2021}, such as those found around the flagellum motor. For simulations using the Oldroyd-B model, we use a small diffusion coefficient $D$ that is proportional to the grid spacing. While it is unclear the degree to which the Giesekus model exhibits similar instabilities, our numerical tests indicate that stress diffusion is not needed for the non-zero values of $\alpha$ used in this study.

The flagellar element is described by a version of Kirchhoff rod theory, in which the flagellum is modeled as a thin rod and stresses are applied on cross-sections along the rod \cite{lim2012}. The configuration of the flagellum is described by the current physical configuration of its centerline, $\bchi\parens{s,t}$, and an orthonormal set of unit director vectors attached to the centerline, $\{\dirVec^1\parens{s,t},\dirVec^2\parens{s,t},\dirVec^3\parens{s,t}\}$, in which $s$ is a Lagrangian parameter with $0\leq s\leq L$, $L$ is the length of the flagellum, and $t$ is the time.

To describe the force and torque balance along the filament, let $\FF^\text{rod}\parens{s,t}$ and $\NN^\text{rod}\parens{s,t}$ be the force and moment, respectively, that are generated across a cross-section of the rod at point $s$ along the center-line. Let $\FF\parens{s,t}$ and $\NN\parens{s,t}$ be the applied force and torque densities of the fluid on the filament. Then the momentum and angular momentum balance equations are
\begin{align}
\mathbf{0} &= \FF\parens{s,t} + \frac{\partial \FF^\text{rod}\parens{s,t}}{\partial s} \label{eq:force},\\
\mathbf{0} &= \NN\parens{s,t} + \frac{\partial \NN^\text{rod}\parens{s,t}}{\partial s} + \parens{\frac{\partial \bchi\parens{s,t}}{\partial s}\times \FF^\text{rod}\parens{s,t}}. \label{eq:torque}
\end{align}
The constitutive relations for the force and moment are derived from a variational argument using the elastic energy potential. In particular, we use a bistable energy formulation that allows for two different stable helices \cite{ko2017}. We expand $\FF^\text{rod}\parens{s,t}$ and $\NN^\text{rod}\parens{s,t}$ in the basis of the local director vectors
\begin{equation}
\FF^\text{rod}\parens{s,t} = \sum_{i = 1}^3 F_i^\text{rod}\dirVec^i\parens{s,t} \text{ and } \NN^\text{rod}\parens{s,t} = \sum_{i = 1}^3 N_i^\text{rod}\dirVec^i\parens{s,t}.
\end{equation}
Following \cite{ko2017}, we utilize a bistable energy formulation that allows for two different stable helices
\begin{align}
E &= E_\text{bend} + E_\text{twist} + E_\text{shear} + E_\text{shear}, \\
E_\text{bend} &= \frac{1}{2}\int_0^L \parens{a_1\left(\Omega_1\parens{s,t}-\kappa_1\right)^2 + a_2\left(\Omega_2\parens{s,t}-\kappa_2\right)^2}\text{d}s,\\
E_\text{twist} &= \int_0^L \parens{\frac{a_3}{4}\left(\Omega_3\parens{s,t}-\tau_1\right)^2\left(\Omega_3\parens{s,t}-\tau_2\right)^2 + \frac{\gamma^2}{2}\left(\frac{\partial\Omega_3\parens{s,t}}{\partial s}\right)^2}\text{d}s,\\
E_\text{shear} &= \frac{1}{2}\int_0^L\parens{b_1\left(\frac{\partial\bchi\parens{s,t}}{\partial s}\cdot\dirVec^1\parens{s,t}\right)^2 + b_2\left(\frac{\partial\bchi\parens{s,t}}{\partial s}\cdot\dirVec^2\parens{s,t}\right)^2}\text{d}s,\\
E_\text{stretch} &= \frac{1}{2}\int_0^L b_3\left(\frac{\partial\bchi\parens{s,t}}{\partial s}\cdot\dirVec^3\parens{s,t}-1\right)^2\text{d}s,
\end{align}
which results in the forces
\begin{align}
F_i^\text{rod} &= b_i\parens{\frac{\partial \bchi\parens{s,t}}{\partial s}\cdot\dirVec^i\parens{s,t} - \delta_{3i}},\quad i=1,2,3, \label{eq:penalty_Force}\\
N_1^\text{rod} &= a_1\parens{\Omega_1\parens{s,t} - \kappa_1}, \\
N_2^\text{rod} &= a_2\parens{\Omega_2\parens{s,t} - \kappa_2}, \\
N_3^\text{rod} &= a_3\parens{\Omega_3\parens{s,t} - \tau_1}\parens{\Omega_3\parens{s,t} - \tau_2}\parens{\Omega_3\parens{s,t} - \frac{\tau_1+\tau_2}{2}} - \gamma^2\frac{\partial^2\Omega_3\parens{s,t}}{\partial s^2}, \label{eq:twist_N}
\end{align}
in which $\delta_{3i}$ is the Kronecker delta function and $\gamma$ is the twist-gradient coefficient. The coefficients $a_1$ and $a_2$ are the bending moduli and $a_3$ is the twist modulus. In standard Kirchhoff rod theory, $\dirVec^3$ is constrained to be the tangential vector of the structure. In the present work, equation \eqref{eq:penalty_Force} with $i = 3$ instead provides a penalty force that approximately enforces this constraint \cite{lim2008a}. The coefficients $b_1$ and $b_2$ are the shear moduli and $b_3$ is the stretching modulus. The strain twist vector $\parens{\Omega_1, \Omega_2, \Omega_3}$, in which $\Omega_i = \frac{\partial \dirVec^j\parens{s,t}}{\partial s}\cdot\dirVec^k\parens{s,t}$ and $(i,j,k)$ is a cyclic permutation of $(1,2,3)$, determines the geometric properties of the helical flagellum. The parameter $\kappa = \sqrt{\kappa_1^2+\kappa_2^2}$ is the intrinsic curvature, and $\tau_1$ and $\tau_2$ are the intrinsic twists, which we assume are constant throughout the simulation. Given the curvature $\hat{\kappa} = \sqrt{\Omega_1^2 + \Omega_2^2}$ and twist $\hat{\tau} = \Omega_3$ of the flagellum, the resulting helix has the radius $R$ and pitch $P$ with
\begin{align}
R = \frac{\hat{\kappa}}{\hat{\kappa}^2+\hat{\tau}^2} \text{ and } P = \frac{2\pi\hat{\tau}}{\hat{\kappa}^2+\hat{\tau}^2}. \label{eq:PitchAndRadius}
\end{align}
Although we do not denote them as such, the values of $\kappa_1$, $\kappa_2$, $\tau_1$, and $\tau_2$ need not be constant along the flagellum, which we exploit to set different properties to the flexible hook that separates the motor from the flagellar filament. Although this work does not consider transition between helical shapes, we retain the bistable energy formulation used in by \cite{ko2017} so that future studies can study the effect of polymorphic transition in viscoelastic fluids.

The force density $\ff\parens{\xx,t}$ is generated by the deformation of the rotating elastic flagellum
\begin{equation}
\ff\parens{\xx,t} = \int_0^L -\FF\parens{s,t}\,\delta_w\parens{\xx-\bchi\parens{s,t}}\,\text{d}s + \frac{1}{2}\grad\times\int_0^L -\NN\parens{s,t}\,\delta_w\parens{\xx-\bchi\parens{s,t}}\,\text{d}s,
\end{equation}
in which $\FF\parens{s,t}$ and $\NN\parens{s,t}$ are the force and torque applied by the fluid on the flagellum and $\delta_w(\xx)$ is a smooth, compactly supported kernel function that mediates coupling between the Lagrangian and Eulerian variables. The linear and angular velocities at the filament are computed by interpolating the Eulerian velocity onto the filament
\begin{align}
\frac{\partial \bchi\parens{s,t}}{\partial t} &= \int_\mathcal{B} \uu\parens{\xx,t}\,\delta_w\parens{\bchi\parens{s,t}-\xx}\,\text{d}\xx, \\
\frac{\partial \dirVec^i\parens{s,t}}{\partial t} &= \frac{1}{2}\int_\mathcal{B} \grad\times\uu\parens{\xx,t}\,\delta_w\parens{\bchi\parens{s,t}-\xx}\,\text{d}\xx,\quad \mbox{for } i=1,2,3,
\end{align}
in which $\mathcal{B}$ is the fixed Eulerian domain. In this work we use a delta function based on the three point B-spline kernel \cite{lee2022}. We remark that the smooth kernel function $\delta_w\parens{\xx}$ appears in both the continuum equations as well as the discretized equations. This is in contrast to traditional IB formulations, in which a smooth kernel function like $\delta_w\parens{\xx}$ appears in the discrete equations of motion but not in the continuum equations \cite{lim2008a}. In conventional IB methods, the width of the smoothed delta is a numerical parameter proportional to the Eulerian mesh width that controls the accuracy of the regularized approximation to the singular delta function. In this model, the width of the delta function is a physical parameter of the model which can be viewed as controlling the effective thickness of the rod. Kernels with larger width yields rods with larger effective thickness.

\begin{table}
\caption{Table of physical and computational parameters for both the model flagellum and the fluid.}
\begin{center}
\begin{tabular}{|c c c|}
\hline
Parameter & Symbol & Value \\
\hline
Helical Radius & $R_0$ & \SI{0.2067}{\micro\metre} \\
Flagellum Length & $L$ & \SI{6}{\micro\metre}\\
Hook length & $L_\text{h}$ & \SI[per-mode=reciprocal]{80}{\nano\meter} \\
Shear modulus & $b_1,b_2$ & \SI{0.8}{\gram.\micro\metre\per\second\squared}\\
Stretch modulus & $b_3$ & \SI{0.8}{\gram.\micro\meter\per\second\squared}\\
Bending modulus & $a_1,a_2$ & \SI{3.5e-3}{\gram.\micro\meter\cubed\per\second\squared}\\
Twist modulus & $a_3$ & \SI{1.0e-4}{\gram.\micro\meter\tothe{5}\per\second\squared}\\
Twist-gradient coefficient & $\gamma$ & \SI{1.0e-3}{\gram\tothe{1/2}.\micro\meter\tothe{3/2}\per\second}\\
Intrinisic curvature & $\kappa$ & \SI[per-mode=reciprocal]{1.3057}{\per\micro\meter}\\
Right-hand intrinsic twist & $\tau_1$ & \SI[per-mode=reciprocal]{-2.1475}{\per\micro\meter}\\
Left-hand intrinsic twist & $\tau_2$ & \SI[per-mode=reciprocal]{2.1475}{\per\micro\meter}\\
Newtonian viscosity & $\mun$ & $\frac{2}{3}\times$\SI{e-6}{\gram\per\micro\meter\per\second}\\
Polymeric viscosity & $\mup$ & $\frac{1}{3}\times$\SI{e-6}{\gram\per\micro\meter\per\second}\\
Fluid density & $\rho$ & \SI{1.0e-12}{\gram\per\micro\meter\cubed}\\
Nonlinearity parameter & $\alpha$ & varies\\
Relaxation time & $\lambda$ & varies \\
Time step size & $\Delta t$ & \SI{2.5e-7}{\second}\\
Filament grid size & $\Delta s$ & \SI{0.04}{\micro\meter}\\
Finest fluid grid size & $\Delta x$ & \SI{0.0391}{\micro\meter}\\
\hline
\end{tabular}
\end{center}
\label{table:parameters}
\end{table}

The initial radius of the helical filament is
\begin{equation}
    R\parens{s} = \left\{\begin{array}{cc}
 0, & 0\leq s\leq L_h, \\
 R_0\parens{1-e^{-c\parens{s-L_h}^2}}, & L_h\leq s\leq L_h+L_f,\end{array}\right.
\end{equation}
in which $c = \SI[per-mode=reciprocal]{2}{\per\micro\meter\squared}$. The helical filament is then initialized as
\begin{equation}
\XX\parens{s,0} = \left[R(s)\cos(2\pi s / \bar{\omega}), R(s)\sin(2\pi s / \bar{\omega}), s\right]\tran
\end{equation}
in which $\bar{\omega} = \SI[per-mode=reciprocal]{0.468}{\micro\meter}$ is the wave number of the helix. The helical radius is \SI{0}{\micro\meter} for the hook and gradually increases to a radius of $R_0$. The vector $\dirVec^3\parens{s,t}$ is initially set as the unit tangent vector to the flagellum, and $\dirVec^1\parens{s,t}$ and $\dirVec^2\parens{s,t}$ are the normal and binormal unit vectors. The flagellum is driven by a rotary motor. We fix the point at $s = 0$ in space and specify the rotation of the director vectors as
\begin{align}
\dirVec^1\parens{0,t} &= \parens{\cos\parens{2\pi\omega t}, -\sin\parens{2\pi\omega t}, 0}, \\
\dirVec^2\parens{0,t} &= \parens{\sin\parens{2\pi\omega t}, \cos\parens{2\pi\omega t}, 0}, \text{ and} \\
\dirVec^3\parens{0,t} &= \parens{0,0,1},
\end{align}
in which $\omega$ is the specified rotation rate. The sign of $\omega$ determines the direction of the rotation. The rotation of the motor generates a torque that is then transmitted to the flagellum through the compliant hook. The length of the hook for an \textit{E. coli} bacterium ranges from \SI{50}{\nano\meter} to \SI{80}{\nano\meter}. Here we specify the hook's length to be $L_\text{h} = \SI{80}{\nano\meter}$. To make the hook flexible, we specify its bending modulus to be two orders of magnitude smaller than that of the filament \cite{son2013,jabbarzadeh2018}. We also assume the hook to be intrinsically straight, so that $\tau = \kappa = \SI[per-mode=reciprocal]{0}{\per\micro\meter}$.

The model is implemented in the open source software IBAMR \cite{ibamr}, which is an MPI parallelized implementation of the immersed boundary (IB) method with adaptive mesh refinement (AMR). The software uses SAMRAI \cite{hornung2002,hornung2006} for adaptive mesh refinement and PETSc \cite{petsc_users_manual,petsc_web_page,balay1997} for iterative linear solvers along with custom preconditioners for AMR discretizations of the incompressible Navier-Stokes equations.

The equations are discretized on a staggered grid in which the normal components of velocities are stored on cell sides and the pressures and components of the conformation tensor are stored on cell centers \cite{harlow1965}. We use second-order finite differences to discretize the Laplacian, velocity gradients, and stress divergence \cite{barrett2019}. The advective derivative is discretized with a second order wave propagation algorithm \cite{ketcheson2013}.

We discretize in time using the implicit trapezoidal rule for the viscous terms and an explicit predictor corrector method for the conformation tensor. The resulting linear system is solved using a GMRES algorithm with a projection method as a preconditioner \cite{griffith2009}. The details of the spatial and temporal discretizations are discussed in previous work \cite{gruninger2024,barrett2019}.

The filament is placed in a computational domain $\mathcal{B}$ which is a cube of length $H = \frac{10}{3}L = \SI{20}{\micro\meter}$. At the physical boundaries, we specify zero velocity and use linear extrapolation for the stress. The computational domain is discretized such that $N = 512$ points in each direction would fill the finest level of the AMR grid. We discretize the Lagrangian structure so that the structure contains approximately one point per Eulerian grid cell.

\section{Results}
\begin{figure}
\begin{center}
    \phantomsubcaption\label{fig:during_sim:superimposed}
    \phantomsubcaption\label{fig:during_sim:stress_dist}
    \includegraphics[width=0.6\columnwidth]{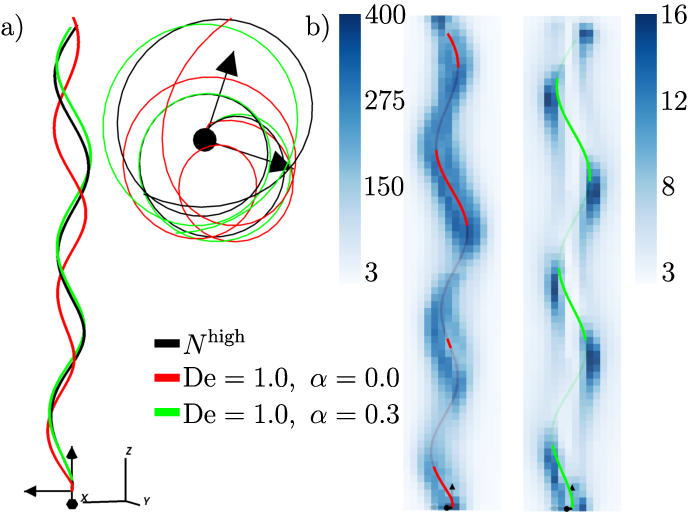}
\end{center}
    \caption{Panel a) shows three superimposed flagella from the side and top perspectives. The black curve corresponds to a flagellum in the fluid $\Nhigh$; the green curve corresponds to a flagellum in a viscoelastic fluid with $\De = 1.0$ and $\alpha = 0.3$, and the red curve corresponds to a flagellum in a viscoelastic fluid with $\De=1.0$ and $\alpha=0.0$. The shapes between the green and black curves are closer than the shape of the red curve. We also show the motor triads at the beginning of the flagellum. The physical location is fixed in place, and the rotation rate of the triads is prescribed. Panel b) shows the trace of the conformation tensor $\CC$ along a slice parallel to the flagellum with $\De = 1.0$ and $\alpha = 0.0$ (left) and $\alpha = 0.3$ (right). Note that for $\alpha = 0.0$, we observe trace values roughly 25 times larger than for $\alpha = 0.3$. Additionally, large regions of stress are found throughout the entire path the flagellum traces for $\alpha = 0.0$, while for $\alpha = 0.3$, the stress quickly dissipates away from the flagellum. Brighter and darker regions of the flagellum highlight the flagellum being in front or behind of the plane.}
    \label{fig:during_sim}
\end{figure}

The flagellum is placed in a viscoelastic fluid, and the motor rotates at a prescribed rate of $\omega = \SI[parse-numbers=false]{2\pi 100}{\radian\per\second}$. We can classify the elasticity of the fluid by using the nondimensional Deborah number, which is the ratio of the polymeric relaxation time $\lambda$ to a characteristic timescale of the structure $T_\text{s}$. We define $T_\text{s}$ to be the time for the motor to complete one turn, so that $T_\text{s} = \SI[parse-numbers=false]{\frac{2\pi}{\omega}}{\second} = \SI{0.01}{\second}$. This gives a Deborah number of $\De = \frac{\lambda}{2\pi/\omega}$. The total viscosity of the fluid is given by $\mu = \mun + \mup$. The viscosities for the viscoelastic fluids are $\mup = \frac{1}{3}\times\SI{e-6}{\gram\per\micro\meter\per\s}$ and $\mun = \frac{2}{3}\times\SI{e-6}{\gram\per\micro\meter\per\s}$, giving a viscosity ratio of $\beta = \frac{\mup}{\mun + \mup} = \frac{1}{3}$. We also simulate a flagellum in two Newtonian fluids with $\mup = 0$: one with $\mun = \SI{1e-6}{\gram\per\micro\meter\per\s}$, which we refer to as $\Nhigh$, and one with $\mun = \frac{2}{3}\times\SI{e-6}{\gram\per\micro\meter\per\s}$, which we refer to as $\Nlow$. The $\Nhigh$ fluid has viscosity matching the total viscosity of the viscoelastic fluids, whereas the $\Nlow$ fluid has viscosity matching only the Newtonian solvent viscosity of the viscoelastic fluids. We define the Reynolds number as $\Re = \frac{\rho U R_0}{\mu}$ in which $U = R_0 / T_\text{s}$ is the characteristic velocity of the system. This yields a Reynolds number of approximately $\Re \approx \SI{4.27e-6}{}$ for $\Nhigh$ and $\Re \approx \SI{6.41e-6}{}$ for $\Nlow$. Note that the Reynolds number for the Giesekus fluid is not well defined, as the effective viscosity of the fluid is shear rate dependent. If we use the zero shear rate viscosity, the Reynolds number for all Giesekus fluids is $\Re\approx \SI{4.27e-6}{}$, that of the $\Nhigh$ fluid. The remaining parameters are given in Table \ref{table:parameters}. For the simulations that follow, we fix the rotation rate and vary the Deborah number $\De$ by the parameter $\lambda$ and the nonlinear parameter $\alpha$ of the polymeric fluid model. Non-zero values of $\alpha$ tune the quadratic nonlinearity of extra stress in the Giesekus model, which controls the degree of shear thinning, and which we demonstrate has profound influences, particularly on the pumping efficiency of the flagellum. In the results that follow, a periodic steady state is achieved after several rotations of the motor. For plots in which average values are shown, we compute the averages after a periodic steady state has been achieved. For fluids with high Deborah numbers and low $\alpha$ values, more cycles are required to achieve a periodic steady state, as the fluid's elastic energy storage capacity is increased. Figure \ref{fig:during_sim:superimposed} shows the shape of superimposed flagella rotating in fluids of varying $\alpha$ values. Figure \ref{fig:during_sim:stress_dist} shows the trace of the conformation tensor around the flagellum. The trace of the conformation tensor is proportional to the elastic strain energy density of the polymeric material \cite{li2017}. We observe trace values that are approximately 25 times larger for the Oldroyd-B fluid ($\alpha = 0.0$) than for the Giesekus fluid with $\alpha = 0.3$. In addition, we observe that large regions of substantial viscoelastic stress exist through the entire path the flagellum traces for $\alpha = 0.0$. For $\alpha = 0.3$, the viscoelastic stresses are localized only near the current location of the flagellum.

\subsection{Domain Dependence}
\begin{figure}
    \centering
    \includegraphics[width=0.75\textwidth]{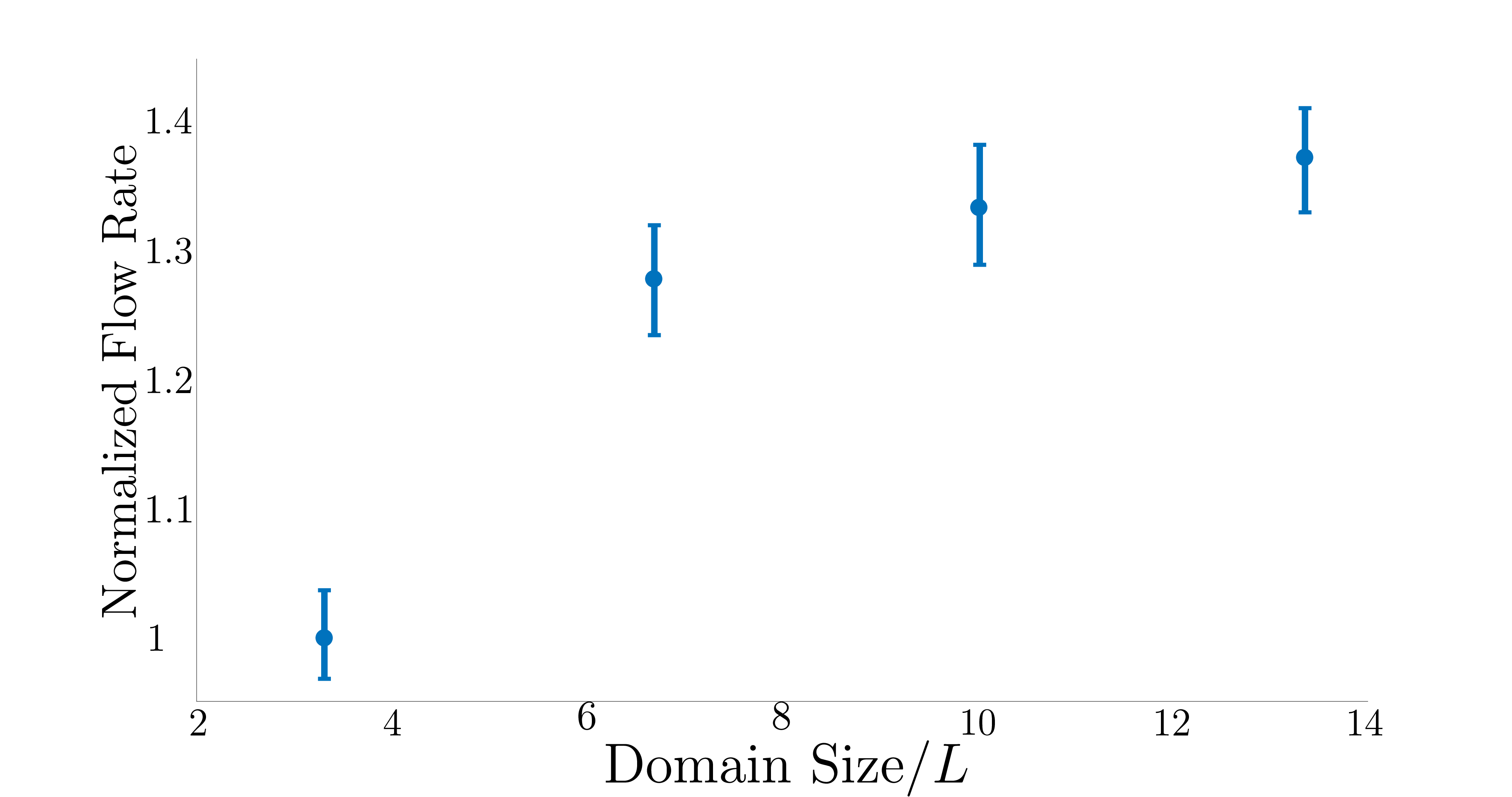}
    \caption{The average flux along with maximum and minimum values through the region $\mathcal{D}$ as a function of the domain size, normalized by the length of the flagellum $L$. We observe an increase in flow rates as the domain size increases. The error bars signify the maximum and minimum flow rates.}
    \label{fig:domain_size}
\end{figure}

Because the model is at very low Reynolds number, the domain can have a profound impact on the results of the simulation. We test the effect that the size of the domain has on the flow by systematically increasing the size of the domain while keeping the grid spacing fixed. We measure the total flux through a circular disk $\mathcal{D}$ above the helix, as will be described in a later section. The flux is shown in Figure \ref{fig:domain_size} for a Newtonian fluid. We observe an increase in flow rates as we increase the size of the domain. We note that due to the hyperbolic nature of the viscoelastic constitutive law, there is little transport of viscoelastic stress towards the boundary, so we expect the boundary effects from viscoelasticity to be comparatively minor. The simulations that follow use a domain size of $L = \SI{10}{\micro\meter}$.

\subsection{Flagellum Shape}
\begin{figure}
    \begin{center}
    \phantomsubcaption\label{fig:shape:radius_mid}
    \phantomsubcaption\label{fig:shape:pitch_mid}
    \phantomsubcaption\label{fig:shape:radius_end}
    \phantomsubcaption\label{fig:shape:pitch_end}
    \includegraphics[width=0.9\textwidth]{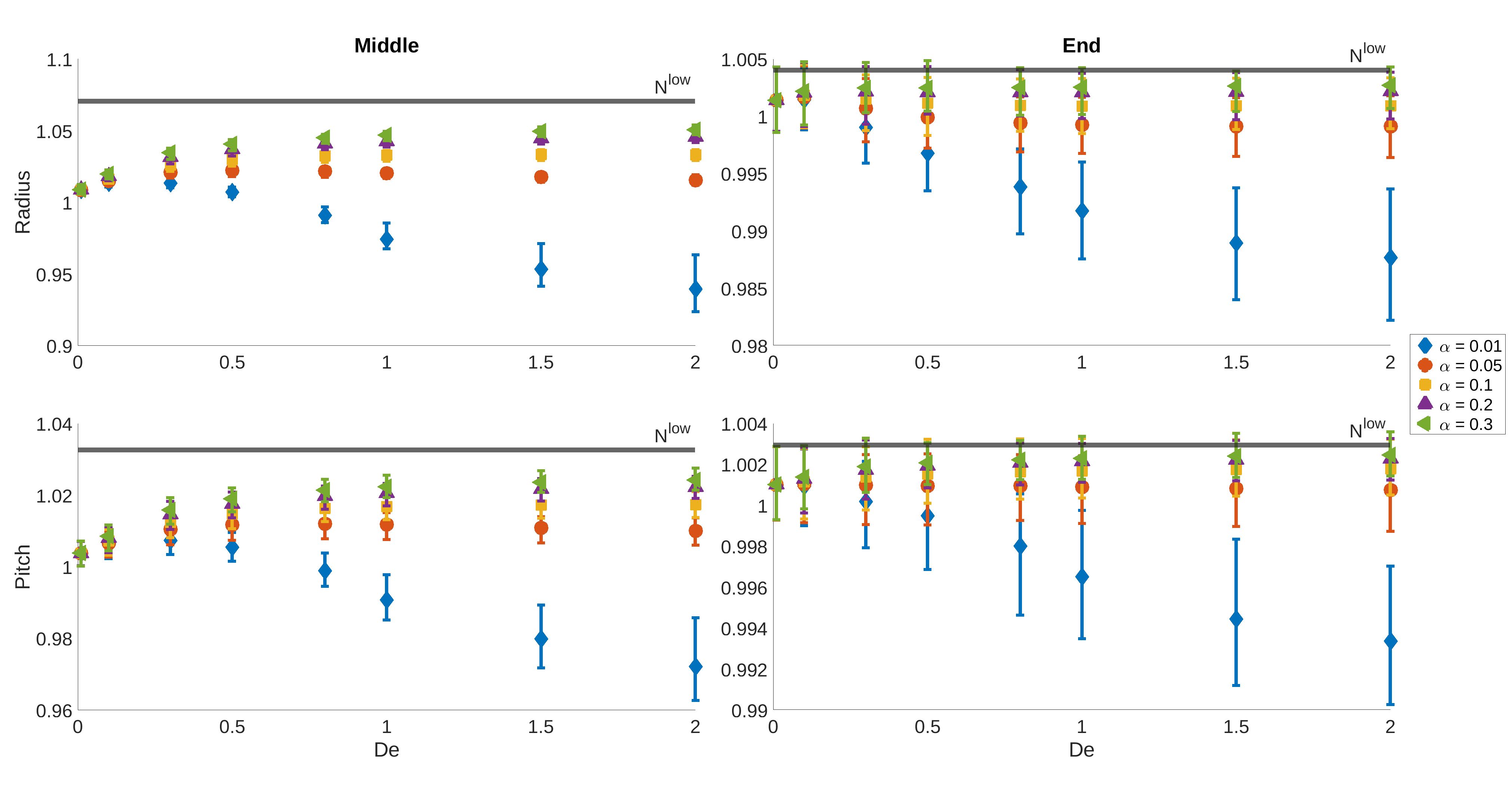}
    \end{center}
    \caption{The time-averaged and maximum and minimum radius (top row) and pitch (bottom row) of the flagellum along the middle (left column) and end (right column), as calculated by equation \ref{eq:PitchAndRadius}. The radius and pitch have been normalized by those in $\Nhigh$. For a small value of the nonlinear parameter $\alpha = 0.01$, we observe initial increases in pitch and radius as we increase the Deborah number $\De$, followed by larger decreases, eventually becoming smaller than that of $\Nhigh$. For all other fluids, we observe increases in pitch and radius as we increase $\De$ approaching those of the $\Nlow$ fluid. Note that in all cases, the changes are no more than 10\% of those in $\Nhigh$ and frequently are only a few percentage points.}
    \label{fig:flagella_shape}
\end{figure}

We assess the shape of the flagellum by computing the radius and pitch of the helix using the strain twist vector in equation \eqref{eq:PitchAndRadius}. The twist of the helix is computed as $\hat{\tau} = \Omega_3$, and the curvature is computed as $\hat{\kappa} = \sqrt{\Omega_1^2 + \Omega_2^2}$. Figure \ref{fig:flagella_shape} shows the time-average, maximum, and minimum values of the radius and pitch along the middle ($s = L/2$) and end ($s = L$) of the flagellum as we vary both the Deborah number $\De$ and the nonlinear parameter $\alpha$. The reported radii and pitches are normalized with respect to the time average radius and pitch for the $\Nhigh$ fluid. Across various Deborah numbers, we find that fluids characterized by values of $\alpha$, which indicate a reduced shear-thinning capacity, tend to result in smaller measurements for both the pitch and radius of the flagelum. Additionally, we have found that significant variations in the shape of the flagella primarily result from lower $\alpha$ values and higher Deborah numbers. Therefore, we infer that major changes in the flagellar helix shape stem from increased elastic responses within the fluid. Conversely, the presence of shear-thinning seems to mitigate these effects.

Overall, we find that the shape of the helix remains relatively consistent in the viscoelastic fluids when compared to Newtonian fluids. The shape changes for all parameters tested are less than 10\% different than in a Newtonian fluid, and typically only vary within a few percentage points. This is in contrast to many undulatory swimmers, which swim with modified shapes in viscoelastic fluids \cite{li2017,fu2008a}. While the shape changes only slightly in viscoelastic fluids, shape changes can have drastic effects on flagellar bundling \cite{lee2018}, which are vital to bacterial locomotion. This effect remains an important area of future study. We also note that the apparent large shape differences in Figure \ref{fig:during_sim:superimposed} are a result of the varying pitch. If we project the helix onto the $y-z$ plane, we obtain a sinusoidal curve, whose amplitude is proportional to the radius of the helix and frequency is proportional to the pitch. Slight changes in the pitch can therefore result in a helix appearing out of phase at different time points.

\subsection{Pumping Efficiency}
\begin{figure}
\begin{center}
\includegraphics[width=0.7\columnwidth]{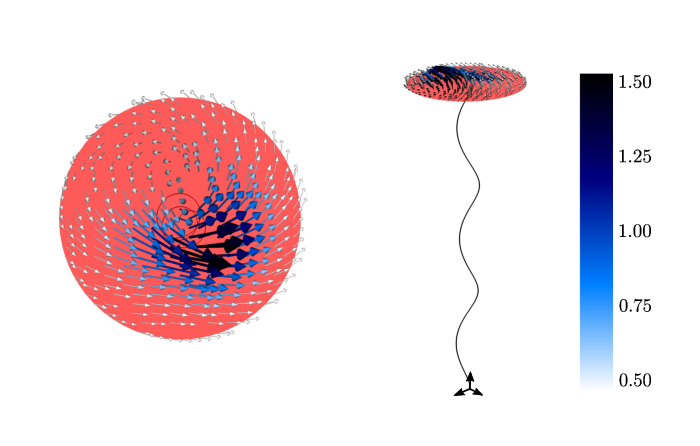}
\end{center}
\caption{The disk $\mathcal{D}$ through which the volumetric flow rate is measured is shown in red. The velocity vectors shown on the region $\mathcal{D}$ correspond to the fluid with $\alpha = 0.3$ and $\De = 1.0$ and have been normalized by the characteristic velocity $U = R_0/T_\text{s} \approx \SI{126}{\micro\meter\per\second}$.}
\label{fig:flux_outline}

\end{figure}
To characterize pumping efficiency, we measure the volumetric flow rate through a disk $\mathcal{D}$ of radius $R = \SI{1}{\micro\meter}$ at the plane $z = \SI{5.5}{\micro\meter}$. Although the flagellum's length is \SI{6}{\micro\meter}, the flagellum's coiled length is approximately \SI{5.25}{\micro\meter}. Consequently, the tip of the flagellum never passes through the disk $\mathcal{D}$. The volumetric flow rate is computed as
\begin{equation}
Q = \int_\mathcal{D} \uu\cdot\nn \,\text{d}A,
\end{equation}
in which $\nn$ is the upward unit normal of the disk $\mathcal{D}$. We note that the choice of the size of the disk $\mathcal{D}$ determines the flux measure, because in the limit as the radius of the disk tends to infinity, the volumetric flow rate $Q$ tends to zero as we encapsulate more regurgitation flow. Figure \ref{fig:flux_outline} shows the region through which the flow rate is calculated. Because we specify the rotation rate of the motor, we can also calculate the torque required to turn the motor at the specified frequency. We calculate the total torque acting on the hook from the motor using equation \eqref{eq:torque}.

\begin{figure}
\begin{center}
\phantomsubcaption\label{fig:torque_fixed_de:torque_De_0.5}
\phantomsubcaption\label{fig:torque_fixed_de:flow_rate_De_0.5}
\phantomsubcaption\label{fig:torque_fixed_de:torque_De_1.0}
\phantomsubcaption\label{fig:torque_fixed_de:flow_rate_De_1.0}
\includegraphics[width=0.85\columnwidth]{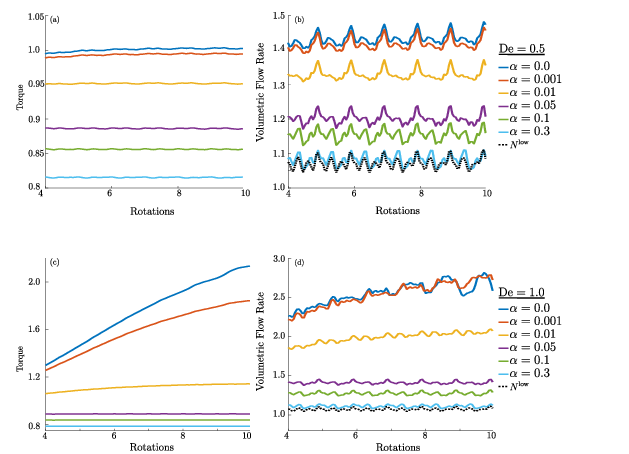}
\end{center}
\caption{Panels (a) and (c) show the torque generated by the motor for varying nonlinear parameter $\alpha$ and fixed Deborah number $\De$ normalized by that in the $\Nhigh$ fluid. We observe a decrease in required torque as we increase the nonlinear parameter $\alpha$. Panels (b) and (d) show the volumetric flow rate through the fixed disk $\mathcal{D}$ normalized by that in the $\Nhigh$ fluid. The top row uses a fixed Deborah number of $\De = 0.5$. The bottom row usese $\De = 1.0$. We observe substantial decreases in the flow rates as the nonlinear parameter increases.}
\label{fig:torque_fixed_de}

\end{figure}

\begin{figure}
    \centering
    \includegraphics[width=0.9\columnwidth]{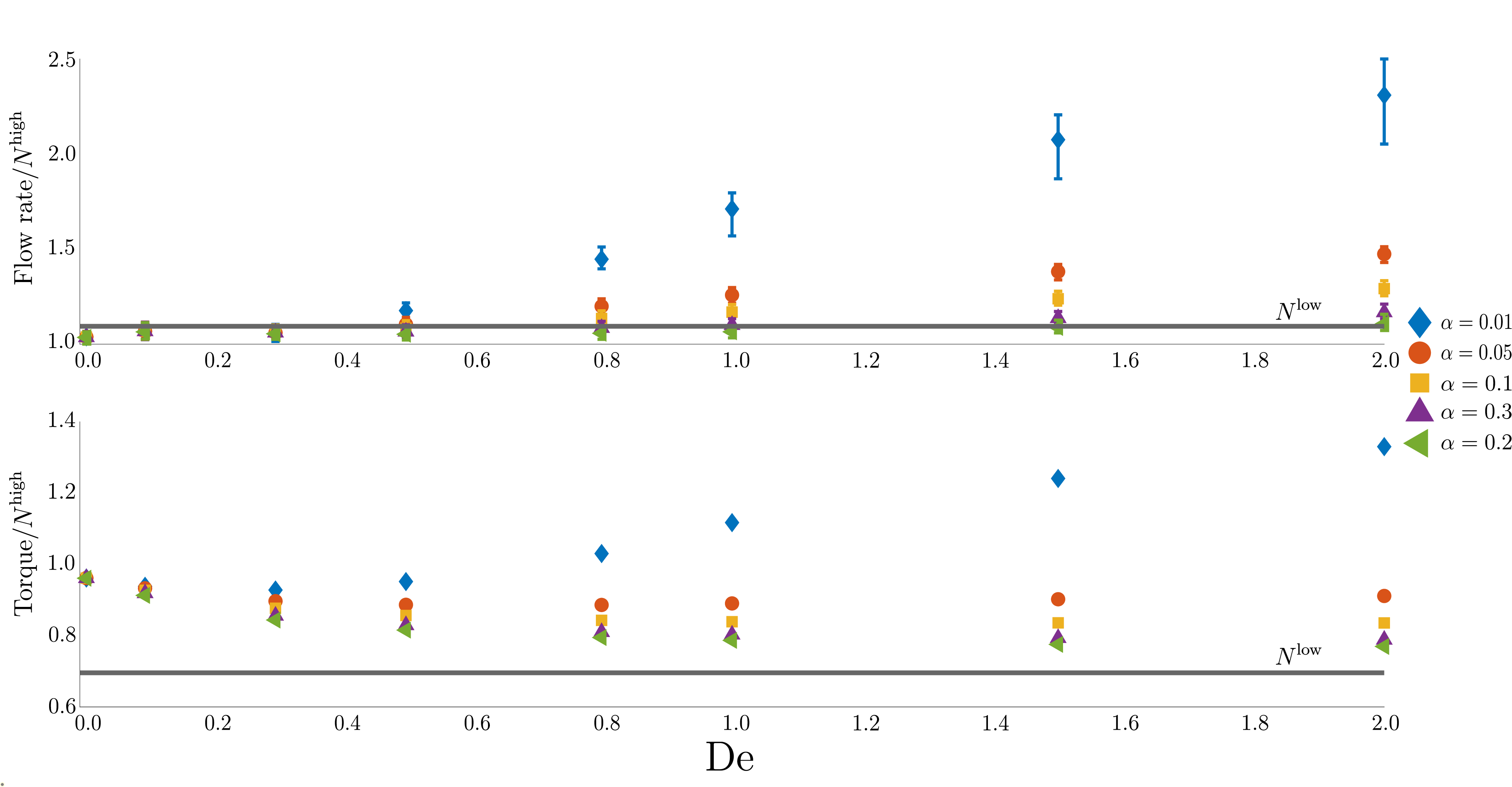}
    \caption{The time-averaged torque and flow rate for various viscoelastic fluids normalized by the torque and flow rate for the $\Nhigh$ fluid. Also shown by the error bars is the maximum and minimum flow rates across a motor rotation. Error bars are not shown for the torque, as the torque is relatively constant in time, see Figure \ref{fig:torque_fixed_de}. We observe that for small $\alpha$ values, the torque and flow rates increase as we increase the Deborah number. For larger $\alpha$ values, the torque decreases as we increase the Deborah number, and the flow rates marginally increase.}
    \label{fig:torque_flow_rate_all}
\end{figure}

Figure \ref{fig:torque_fixed_de} shows the flow rate $Q$ through the region $\mathcal{D}$ as well as the total torque acting on the hook for two fixed Deborah numbers as we vary the nonlinear parameter $\alpha$. We find that a steady state is quickly reached after the first few rotations of the motor. The exception is for larger Deborah numbers and small $\alpha$ values, which require additional rotations to reach a steady state. This is indicative of the instabilities inherent in the Oldroyd-B model for very high Deborah numbers \cite{thomases2014}; although, because of the inclusion of stress diffusion, we do anticipate that steady state values will eventually be achieved for all Deborah numbers. Figure \ref{fig:torque_flow_rate_all} shows the time-averaged, maximum, and minimum flow rate and time-averaged torque for each fluid, normalized with respect to the time averaged values obtained for the $\Nhigh$ fluid. For a small nonlinear parameter value of $\alpha = 0.01$, we observe a non-monotonic required torque to spin the motor as a function of the Deborah number. The required torque initially decreases, achieving a minimum at approximately $\De = 0.3$, followed by increases as we increase the Deborah number further. For Deborah number $\De = 2.0$, the torque required to spin the motor is nearly 1.4 times that of the required torque in the $\Nhigh$ fluid. For all other values of $\alpha$, we observe slow decreases in the required torque as we increase the Deborah number, with the lowest required torque found with $\alpha = 0.3$. For all fluids tested, the required torques were higher than that of flagellum in the $\Nlow$ fluid. This suggests that for large enough $\alpha$ values, the shear-thinning behavior of the viscoelastic fluid reduces the impact of its elasticity, lowering the effective viscosity near the motor below that of the $\Nhigh$ fluid and towards that of the $\Nlow$ fluid.

The flow rate for every fluid considered is larger than that of the $\Nhigh$ fluid. Generally, we observe increases in flow rates as we increase the Deborah number or decrease the nonlinear parameter $\alpha$  with the largest flow rates observed corresponding to the largest De number and smallest $\alpha$ values tested. Similar to the shape of the flagellum, the flow rate varies across a complete rotation of the motor. Further, congruent with our findings regarding the shape variation of the flagellum, we find that flow variations increase as the Deborah number is increased and the nonlinear parameter $\alpha$ is decreased. 

\begin{figure}
\begin{center}
    \includegraphics[width=0.8\textwidth]{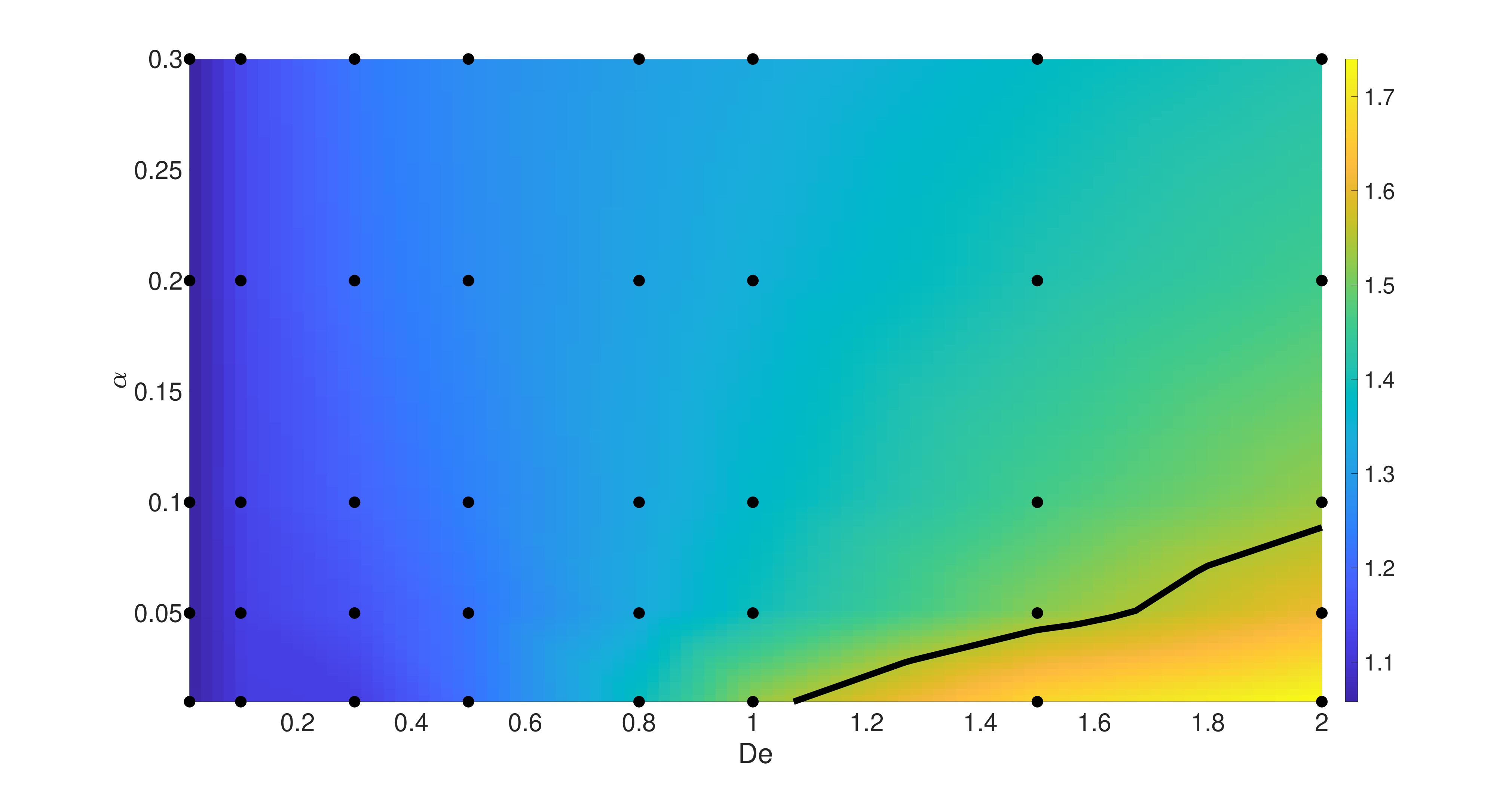}
\end{center}
    \caption{The efficiency of the motor as computed by equation \eqref{eq:eff} as we vary the nonlinear parameter $\alpha$ and the Deborah number $\De$. The black circles are data points from the simulation, and linear interpolation is used to compute the remaining efficiencies. The black curve is the contour on which the efficiency is equal to that of $\Nlow$. For every simulation, we observe a greater efficiency than in the $\Nhigh$ fluid. For small $\alpha$ and large $\De$ values, we observe a greater efficiency than in the $\Nlow$ fluid.}
    \label{fig:eff}
\end{figure}

To assess the performance of the motor, we compute its instantaneous efficiency,
\begin{equation}\label{eq:eff}
    E_i(t) = \frac{Q_i(t) / \tau_i(t)}{Q_{\Nhigh}(t)/\tau_{\Nhigh}(t)},
\end{equation}
in which $\tau_{\Nhigh}(t)$ and $Q_{\Nhigh}(t)$ are the torque and flow rate from the $\Nhigh$ fluid and $\tau_i(t)$ and $Q_i(t)$ are the torque and flow rate from fluid $i$. We then average the efficiency $E_i(t)$ over the last five rotations of the motor. Efficiency values greater than one imply that the motor in fluid $i$ is able to pump fluid more efficiently than the motor in fluid $\Nhigh$. Figure \ref{fig:eff} shows the efficiency for all fluids tested. We perform linear interpolation of the efficiency value for combinations of $\De$ and $\alpha$ that were not simulated. In all simulations, the efficiency of the motor is greater than the efficiency in the $\Nhigh$ fluid. The solid black curve in Figure \ref{fig:eff} shows the contour on which the efficiency is equal to that of $\Nlow$. We observe that for small $\alpha$ values and large $\De$ values, the efficiency of the motor is greater than that for $\Nlow$. 

Our findings reveal a nuanced interplay between the viscoelastic fluid's elasticity and shear-thinning properties and their impact on motor efficiency. Increased elasticity necessitates a higher torque to spin the motor, yet it boosts the flow rate to a greater extent, resulting in enhanced motor efficiencies at higher Deborah numbers. Conversely, while shear thinning reduces the required torque to spin the motor, it substantially reduces the flow rate, thereby diminishing motor efficiency at larger values of $\alpha$. Additionally, the enhancement in motor efficiency driven by increasing Deborah number values is more pronounced at lower values of $\alpha$.

These results appear to be inconsistent with recent results on locomotion in complex fluids. In generalized Newtonian fluids that exhibit shear-thinning with negligible elasticity, several groups \cite{qu2020,gomez2017,gagnon2014,hyakutake2019} have demonstrated that organisms swim faster than in comparable Newtonian fluids. The common explanation is that swimmers benefit from a soft confinement effect. The fluid exhibits a lower viscosity in the immediate vicinity of the swimmer, allowing the swimmer to ``push off" the more viscous fluid layer. For fluids with significant elastic properties, a prevailing theory is that swimmers gain a speed boost from the stored energy in the fluid \cite{teran2010,li2017,thomases2014}. For Deborah numbers $\De \geq 1$, the relaxation time of the fluid exceeds the rotation period of the flagellum, meaning the flagellum continues to return to a volume of stored elastic stress. This provides the flagellum with elastic resistance allowing the flagellum to swim faster, analogous to a pusher.

\begin{figure}
    \centering
    \includegraphics[width=0.6\textwidth]{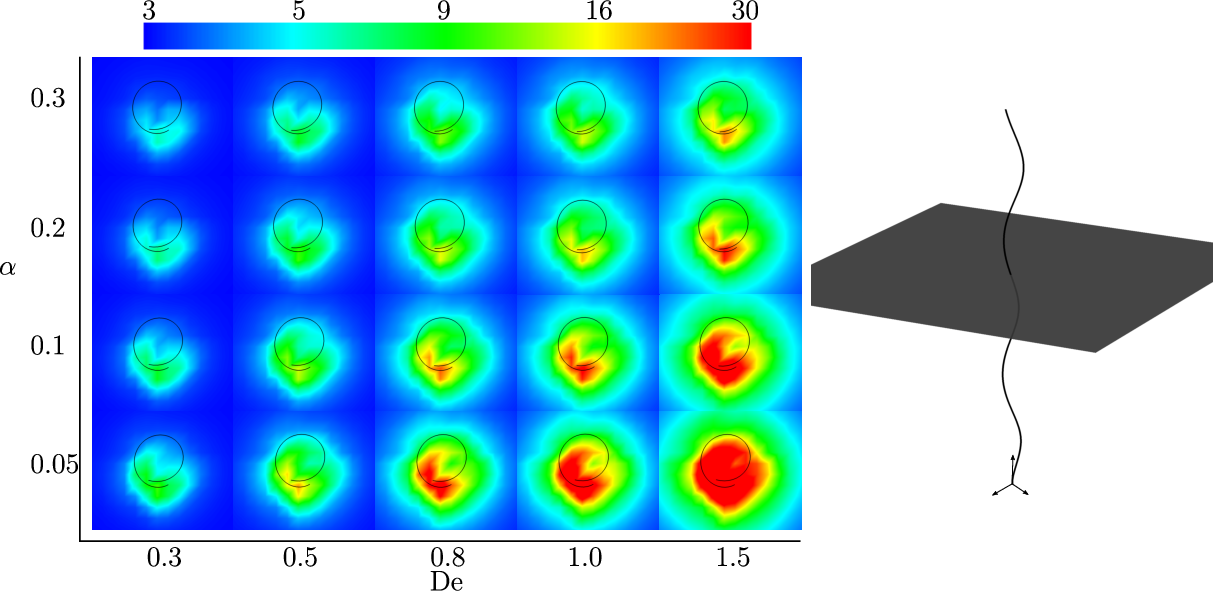}
    \caption{The trace of the conformation tensor $\CC\parens{\xx,t}$ as we vary the Deborah number and the nonlinear parameter. The trace is shown on a slice along the midpoint of the flagellum. We observe that as the Deborah number increases, the flagellum begins to return to regions of stored elastic energy, which enhances the pumping capacity of the flagellum. As we increase the nonlinear parameter, the magnitude of the stored stress sharply decreases, which reduces the effect of fluid elasticity on the flagellum. Note the colorbar uses a logarithmic scaling.}
    \label{fig:tr}
\end{figure}

For the viscoelastic model used here, however, these two stories are at odds with each other. Shear thinning is induced in a neighborhood around the flagella, providing less viscous resistance to the flagellum's motion. Increased shear-thinning requires less torque to spin the motor and therefore weaker elastic resistance for pushing. The consequence of this can be observed in figure \ref{fig:eff}, which shows that the efficiency of the motor decreases as we increase the shear-thinning capacity of the fluid. The elastic energy of the fluid is given by the trace of the conformation tensor \cite{li2017}. Figure \ref{fig:tr} shows a plot of the trace of the conformation tensor along the middle of the flagellum as we vary $\alpha$ and $\De$. As expected, for $\De \geq 1.0$, the elastic energy of the fluid remains high throughout the path that the flagellum traces out in the fluid. However, the magnitude of the trace drastically decreases as we increase $\alpha$. For large $\alpha$ values, the flagellum is returning to areas with lower elastic energies.

\begin{figure}
    \centering
    \includegraphics[width=0.6\textwidth]{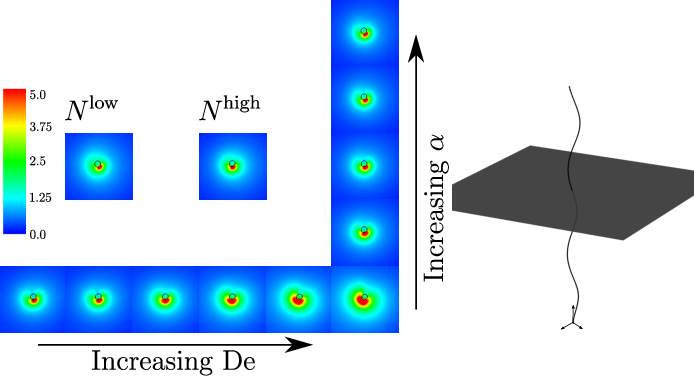}
    \caption{The velocity magnitude as we vary the Deborah number $\De$ and the nonlinear parameter $\alpha$. The velocity magnitude is shown on a slice along the midpoint of the flagellum and has been normalized by the non-dimensional velocity $U = R_0 / T_\text{s}$. As we increase the Deborah number, we observe that the size of the region of non-zero velocities increases. However, as we increase $\alpha$, the size of the region decreases. Also shown are the velocity magnitudes for the Newtonian $\Nlow$ and $\Nhigh$ fluids. As we increase $\alpha$, the size of the region approaches that of the Newtonian fluids, suggesting marginal confinement effects.}
    \label{fig:umag}
\end{figure}

We also do not see a clear confinement effect around the flagellum. Figure \ref{fig:umag} shows the magnitude of the velocity as we vary $\De$ and $\alpha$. For a fixed small nonlinear parameter of $\alpha = 0.01$, we actually observe that the region that experiences non-zero flow increases in size as we increase the Deborah number: the opposite of a confinement effect. As we increase the shear-thinning capacity of the fluid, the size of this region decreases, approaching a size that is comparable to that of the $\Nhigh$ and $\Nlow$ fluid. Because the shear-thinning is limited to the polymeric stress, we do not observe a confinement effect overall, as the effect from the Newtonian solvent becomes dominant.

\section{Discussion and Concluding Remarks}
We have studied both the shape of the flagellum and the motor efficiency in viscoelastic fluids with varying capacities of shear-thinning. We compared these flagella with those in Newtonian fluids, one with viscosity equaling the total of the polymeric and solvent viscosities $\Nhigh$, and one with viscosity equaling that of only the solvent $\Nlow$. We have observed different helical shapes assumed by the flagellum, which for the range of parameters considered in this study, differed by less than 10\% when compared to a Newtonian fluid. While it is unclear how much effect this slight change has on the pumping efficiency of the flagellum, the bundling ability of multiple flagella could be significantly affected and is worth further investigation \cite{lee2018}.

We note that the flagellum elasticity model utilized in this study is based on physical characterizations of the flagella of \textit{E.\ Coli} \cite{goldstein2000,darnton2007a,ko2017}, and this model has been successfully used to study flagellar bundling and locomotion in Newtonian fluids \cite{lim2012,ko2017,jabbarzadeh2018}. The elasticity model has several timescales related to stretching, twisting, and bending, and the interplay between the timescales of the flagellum and the elastic time scale of the fluid is not fully explored in this manuscript. The timescales of the flagellum are derived in the supplementary material. For undulatory swimmers, \cite{thomases2017} found that swimmers can achieve greater speed boosts if their gait can adjust to the elasticity of the fluid. To the authors' knowledge, a similar analysis has not been performed for helical swimmers. For the results discussed herein, when compared to the rotation period of the motor, we find the timescales of the flagellum are much faster than that of the motor, making the flagellum a stiff body. The timescales of the flagellum are also faster than that of the elasticity of the fluid, and the flagellum's shape is not substantially changed by fluid elasticity. However, shear-thinning of the fluid reduces the elastic energy stored by the fluid, which complicates the conclusion reached by \cite{thomases2017}.

The importance of shear-thinning behavior for speed enhancement in helical swimmers has been demonstrated previously. \cite{gomez2017} and \cite{demir2020} both studied helical swimming in shear-thinning fluids with negligible viscoelasticity. They both demonstrated that shear-thinning can greatly enhance swimming speed for helical swimmers, and they concluded that confinement effects are the primary cause of speed enhancement. Demir \etal\ further hypothesized that the shear-thinning effects dominate any viscoelastic effects, due to the magnitude of the swimming enhancement observed in their studies. \cite{qu2020}, in their studies of helical swimming in viscoelastic fluids with shear-thinning, also found that the magnitude of shear-thinning effects are dominant over the viscoelastic effects.

We find that the motor efficiency in all viscoelastic fluids is greater than the efficiency in the $\Nhigh$ fluid. For fluids with large Deborah numbers and small values of the non-linear parameter $\alpha$, the efficiency is greater than the $\Nlow$ fluid. As we increase $\alpha$, the efficiency of the motor decreases. This decrease in motor efficiency can be explained by the complex interplay between the elastic and shear-thinning responses of the viscoleastic fluid. Although greater elasticity enhances the storage of elastic energy in the fluid, facilitating higher swimming speeds and thus increased flow rates \cite{li2017,li2021,spagnolie2013}, shear-thinning has a contrasting effect. It diminishes the torque required for the flagella to rotate, which results in a decreased amount of elastic energy being stored in the surrounding fluid and decreased flow rates. Further, in our results, we do not observe any confinement effect, and the velocity profile approaches that of a purely Newtonian fluid as $\alpha$ is increased. In fact, for small values of $\alpha$, we observe that the region of substantial flow around the flagellum increases in size as we increase the Deborah number. This finding is in contrast to \cite{qu2020}, and this difference in results can be partially attributed to the low Deborah numbers utilized in Qu and Breuer's experiments. Our findings suggest that the Deborah number needs to be larger than unity to fully observe speed enhancements.

Our results suggest that too much shear-thinning in the Giesekus model actually reduces the efficiency of the swimmer. We note that this is not in contradiction with other results. Instead, we claim that the mechanism for shear-thinning is important. For the Giesekus model, the solvent behaves as a Newtonian fluid. If the viscoelastic contribution to the stress experiences excessive shear thinning, the viscoelastic stresses become subdominant to the Newtonian solvent's stresses, making the fluid appear Newtonian. Further, when making comparisons to Newtonian fluids, it is important to carefully select the appropriate Newtonian fluid with which to compare. The Reynolds number measures the relative importance of inertial and shear forces. The classical definition relies on a choice of viscosity, which for shear thinning fluids is shear-rate dependent, which varies in both space and time in our context. In the results considered here, we compare with Newtonian fluids whose Reynolds numbers match with the zero shear rate viscosity of the shear-thinning fluids. As discussed by \cite{thompson2021}, more appropriate comparisons would need to determine the characteristic viscosity near the flagellum. An important step to fully elucidating the mechanisms behind non-Newtonian swimming is to determine the detailed rheology of the fluids in which bacteria swim, and from this, determine an accurate model that fully captures the physical mechanisms of shear-thinning and viscoelasticity. This can be a complicated undertaking, however, because biological fluids typically have multiple relaxation times as well as a multitude of other properties, such as polymer chain entanglements of mucus in the respiratory, gut, and cervicovaginal tracts.

\section*{Acknowledgements}
The authors thank Robert Guy for helpful conversations about motility in viscoelastic fluids. We acknowledge support from NSF awards DMS 1929298, DMS 1853591, DMS 1410873, DMS 1664645, OAC 1450327, OAC 1652541, and OAC 1931516 and NIH awards U01HL143336 and R01HL157631. M.G.F. acknowledges support by the Alfred P. Sloan Foundation award G-2021-14197. C.G. is grateful for support from the Department of Defense (DoD) through the National Defense Science and Engineering Graduate (NDSEG) Fellowship Program. S.L. acknowledges support from the Charles Phelps Taft Research Center, University of Cincinnati.

\printbibliography

\end{document}